# Computational search for materials having a giant anomalous Hall effect in the pyrochlore and spinel crystal structures

Sean Sullivan[1^], Seungjun Lee[2^], Nathan J. Szymanski[1], Amil Merchant[3], Ekin Dogus Cubuk[4], Tony Low[2*], and Christopher J. Bartel[1*]

[^]These authors contributed equally
[*]Correspondence to tlow@umn.edu or cbartel@umn.edu

[1]Department of Chemical Engineering and Materials Science, University of Minnesota, Minneapolis, Minnesota 55455, USA

[2]Department of Electrical Engineering and Computer Engineering, University of Minnesota, Minneapolis, Minnesota 55455, USA

[3]Google DeepMind

[4]No current affiliation

## ABSTRACT

Ferromagnetic pyrochlore and spinel materials with topological flat bands are of interest for their potential to exhibit a giant anomalous Hall effect (AHE). In this work, we present computational predictions of stability and electronic structure for 448 compositions within the pyrochlore ($A_2B_2O_7$) and spinel ($AB_2O_4$) frameworks. Of these, 92 are predicted to be thermodynamically stable or close (< 100 meV/atom) to the convex hull, with trends deviating from expectations based on ionic radius-ratio rules. Thirteen are predicted to adopt a ferromagnetic ground state among the collinear configurations considered. Two additional materials meeting these criteria were also identified from open materials databases. Calculations of anomalous Hall angles (AHA) and conductivities reveal that 11 of the screened materials are promising candidates for spintronic applications requiring high electronic conductivity and a giant AHE. Our results suggest that the AHA can be further enhanced by tuning the Fermi level, for example through chemical doping. Using this approach, we identify five materials whose AHA exceed 0.2 under the approximation of collinear magnetism. Notably, $Ag_2Pt_2O_7$ exhibits a high AHA of 0.405 when its Fermi level is optimized. These findings provide a roadmap for the targeted synthesis of new pyrochlore and spinel compounds with enhanced AHE properties. They also broaden the compositional design space for these structures and support the discovery of high-performance materials for next-generation spintronic applications.

## INTRODUCTION

The anomalous Hall effect (AHE), which is the appearance of a transverse Hall current in the absence of an applied magnetic field, originates from time-reversal symmetry breaking in both ferromagnets and noncollinear antiferromagnets.[1,2] Materials exhibiting the AHE can be utilized in a range of applications including magnetic sensors, memory applications with high sensitivity and good thermal stability[3], and spintronic devices.[4–6] Research in this area has also helped unravel correlations between ferromagnetism, spin-orbit coupling, and spin chirality.[7–9] Nevertheless, it remains difficult to predict which materials will exhibit sizeable AHE. To better understand the AHE and discover new materials that exhibit giant anomalous Hall conductivity (AHC), model systems containing Kagome lattices (e.g., $Co_3Sn_2S_2$[10] and $LiMn_6Sn_6$[11]) or sublattices (e.g., spinel or pyrochlore ferromagnets[12,13]) have been explored.

Materials in the α-pyrochlore (hereafter referred to as pyrochlore) and cubic spinel structures have attracted substantial attention for their application in magnetic devices, electronics, and quantum computing.[14–22] The pyrochlore crystal structure, which takes the general formula $A_2B_2O_7$, is composed of *A* and *B* sites that form two interlinked tetrahedral sublattices. Materials in the pyrochlore structure commonly possess a trivalent *A* cation coordinated to eight O ligands and a tetravalent *B* cation coordinated to six O ligands, although $A^{2+}/B^{5+}$ compositions have also been observed. In contrast, spinel-structured oxides take the formula $AB_2O_4$, where the *A* sites usually contain divalent cations coordinated to four O ligands, and *B* sites contain trivalent cations coordinated to six O ligands. Much like the pyrochlore structure, the octahedral *B* sites in spinel form a corner-sharing network of tetrahedral polyhedra. These networks are illustrated by the conventional unit cells displayed in **Fig. 1**. Along the [111] crystallographic plane, both pyrochlores and spinels resemble the Kagome lattice and therefore host double-degenerate topological flat bands. This degeneracy can be lifted when ferromagnetic ordering leads to broken time-reversal symmetry, resulting in giant AHC even when spin-orbit coupling is relatively weak.[12] Despite the theoretical promise of such materials, there are relatively few experimentally verified spinel and pyrochlore compounds with ferromagnetic ground states, leaving a substantial gap for discovering new AHE materials.

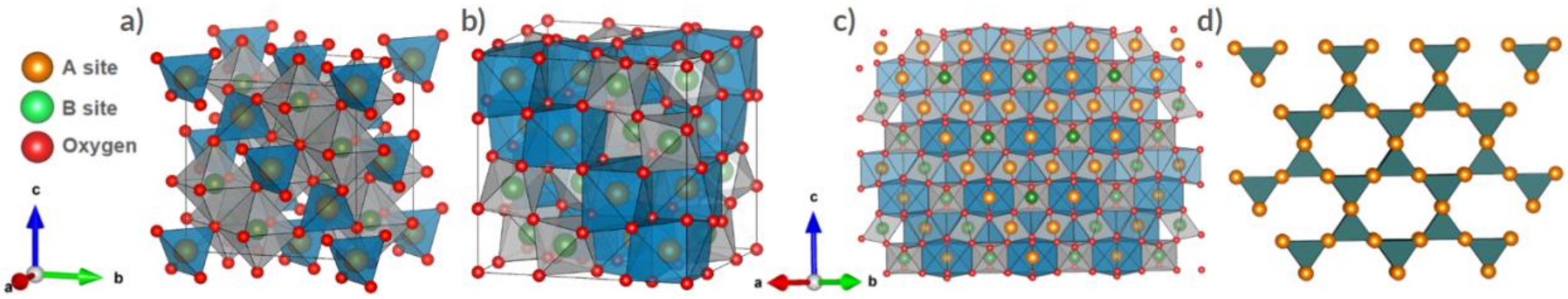

**Figure 1.** Conventional unit cells of the spinel (a) and pyrochlore (b) crystal structures, shown oriented relative to the leftmost axes. (c) Pyrochlore structure as viewed normal to the [111] plane. (d) Kagome-like lattice formed between *A*-site cations.

While there is growing interest in pyrochlore and spinel materials with unique magnetic properties, the discovery of new compounds in each space has been largely driven by empirical rules. For example, the ratio of ionic radii ($r_A/r_B$) between the $A$ and $B$ cations has been used to predict which compounds are likely to crystallize in the pyrochlore structure. Ratios satisfying $1.36 \leq r_A/r_B \leq 1.71$[23] have been suggested for materials of the $A^{3+}/B^{4+}$ type, while larger ratios up to 2.2 have been observed in pyrochlores of the $A^{2+}/B^{5+}$ type[24]. Even larger ratios (up to 2.3) have been observed when using high-pressure synthesis.[25] Rules based on ionic radii have shown some degree of predictive accuracy when the $A$ cation is a rare earth metal and the $B$ cation is a $5d$ transition metal. However, their generality to more diverse cations has not yet been demonstrated, and these rules are likely to change when high-pressure synthesis techniques are employed.[25] While there has been less work on predicting the stability of spinel structures from ionic radii alone, Song and Liu proposed the following tolerance factor[26]:

$$t = \frac{\sqrt{3}(r_B + r_X)}{2(r_A + r_X)}$$

where $r_X$ is the anion radius ($r_O$ in the case of oxides). A composition is then predicted to be stable in the spinel structure if $t < 1.224$. These geometrically inspired rules can help rule out unlikely ground states, but they are also prone to false positives.[27] Having the proper ratio of ionic radii is an insufficient condition to guarantee that a composition will adopt either a spinel or pyrochlore structure, and there have been many observations of materials that satisfy the proposed tolerance factor but whose ground state does not match the predicted structure.

In this work, we explore the vast chemical space of pyrochlore and spinel materials with the goal of identifying novel compounds with giant AHC. Using high-throughput density functional theory (DFT) calculations, we screened for compositions that are thermodynamically stable or reasonably close to the convex hull in the spinel or pyrochlore structure. Magnetic properties and electronic structures were then computed for these stable (or nearly stable) materials, yielding new candidates for giant AHC that were further assessed using Hall conductivity simulations. This approach resulted in the identification of 11 new compounds predicted to adopt the pyrochlore structure and exhibit large AHC and anomalous Hall angle (AHA).

## RESULTS AND DISCUSSION

This work focuses on materials in the pyrochlore and spinel structures (space group $Fd\bar{3}m$) to find new compounds that may exhibit a giant AHE based on the tight binding model introduced in our previous work.[12] Our screening process is schematically illustrated in **Fig. 2**. As a starting point, we searched the Materials Project[28] database for all materials with a composition of $A_2B_2O_7$ (pyrochlore) or $AB_2O_4$ (spinel), and a space group of $Fd\bar{3}m$. These entries were filtered to only include those that are thermodynamically stable. From a total of about 150,000 entries in Materials Project, we found 72 pyrochlore and 34 spinel materials that satisfy these criteria. Of these entries, 42 pyrochlores and 26 spinels are associated with entries in the Inorganic Crystal Structure Database (ICSD)[29], suggesting they have been experimentally synthesized, and 40 of them are reported to have a ferromagnetic ground state in the Materials Project. A separate investigation of stable compounds in the GNoME dataset[30] reveals 54 additional materials with spinel or pyrochlore ground-states that are on the convex hull. From the combined set of 160 materials sourced from Materials Project and GNoME, many contain lanthanides and actinides. To narrow the scope of our investigation and because of challenges in accurately modeling *f*-electron systems with DFT, we considered only those containing $La^{3+}$, $Ce^{3+}$, or $Pr^{3+}$ in this work. This filtering eliminated 73 materials from our study, leaving 87 compounds – 61 from the Materials Project and 16 from GNoME – as candidates for further calculations.

To expand the library of candidate materials in the spinel and pyrochlore structures, we used a data-mined structure prediction algorithm (DMSP)[31] to generate new hypothetical structures. This algorithm yields a probability that ionic substitution into a given structure will produce a thermodynamically stable material (see **Methods** for details on the application of DMSP). Because we are interested in finding materials with ferromagnetic ground states, we only considered ionic substitutions that are likely to introduce unpaired $d$-electrons. Since the $A$ and $B$ sites are coordinated with oxygen in each structure, we assumed that high-spin configurations will be adopted. Starting from the 72 pyrochlore and 34 spinel structures in Materials Project, this substitution process resulted in 654 compounds, of which 448 of these compounds are predicted by DMSP to be high-probability substitutions (338 pyrochlores and 110 spinels). DFT calculations were then performed on these 448 candidate materials to optimize their geometries and obtain formation energies consistent with the Materials Project – that is, at the PBE(+U) level of theory with the appropriate corrections.[32] The thermodynamic stability of each ion-substituted compound was evaluated by comparing these formation energies to all relevant competing phases in the Materials Project database using the convex hull formalism.[33] Initial high-throughput calculations performed using PBE(+U) found that 10 of these 448 compounds (10 pyrochlores, 0 spinels) are thermodynamically stable, and an additional 56 compounds (42 pyrochlores, 14 spinels) are within 100 meV/atom of the hull. These stable (or nearly stable) compounds were added to the list of 61 compounds from the Materials Project and 16 compounds from GNoME, resulting in 143 candidates that were subject to further calculations. By including materials above the hull, we account for the fact that some may be computed to be metastable at 0 K but still synthetically accessible under some conditions.[33,34] Details on all the materials generated through DMSP, including those that relaxed to non-$Fd\bar{3}m$ space groups and therefore excluded from later analysis, are provided in **Table S1**.[35]

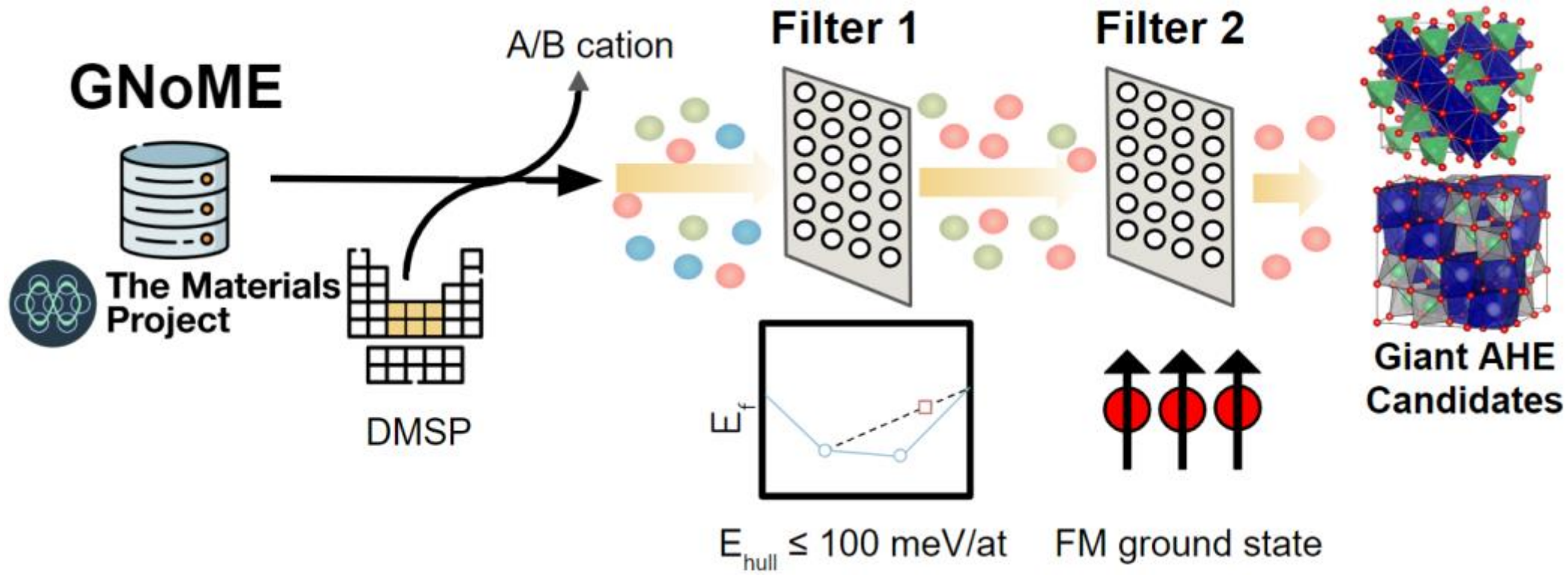

**Figure 2.** An overview of the candidate screening process. Compounds from the Materials Project and GNoME were subjected to ionic substitutions with likely magnetic cations, and the stability (first filter) of each composition was evaluated in the pyrochlore and spinel structure prototypes. Materials calculated with DFT to be stable or close to the convex hull ($E_{hull}$ < 100 meV/atom) were then passed through a second filter including magnetic initialization sampling. Only those with ferromagnetic ordering are considered as potential AHE candidates.

Before performing more involved calculations related to the AHE, we first explored the trends in thermodynamic stability across the pyrochlore and spinel materials. In **Fig. 3**, we map the stability of 446 compositions in the pyrochlore structures: 114 from the Materials Project, 315 generated with DMSP, and 17 from GNoME. Of these materials, 46 are reported in the ICSD as having the pyrochlore structure (indicated with a ▽ in **Fig. 3**), 21 of which are calculated to be thermodynamically stable. Another 21 of these ICSD materials are unstable but within 100 meV/atom of the hull. Through DMSP, our search for new pyrochlore materials adds another 52 compounds within 100 meV/atom of the hull. In total, 138 compounds in **Fig 3.** are predicted to lie within this potentially accessible range, 42 having ICSD entries and 96 that, to our knowledge, have not been synthesized. We note that the 100 meV/atom cutoff for experimental accessibility is not a strict boundary but a practical threshold consistent with prior work.[34] Some compounds above this value may still be stabilized under specific synthesis conditions, and it may be possible that compounds within this boundary are not synthetically accessible if there are no conditions under which they can be preferentially stabilized. More work is needed to determine the synthetic accessibility of these newly predicted compounds, but their thermodynamic stability or relatively small instability suggests they are promising candidates for synthesis. These computed energies also depend on the library of known materials in each chemical space. For the current analysis, we only used compounds from the Materials Project and GNoME dataset to assess thermodynamic stability.

While the rare-earth pyrochlores have been well studied, our search expands the space of known materials to include new combinations of elements on the *A* and *B* sites. Although most transition metals are highly unstable in the *A* site of the pyrochlore structure, there are some exceptions including two early transition metals (Sc and Y) and several late transition metals (Cu, Ag, Au, Cd, and Hg). We suspect this observation can be related to the electron configuration of each metal and its stability in the 8-coordinated *A* site of the pyrochlore structure. While the intermediate transition metals have a strong preference to adopt octahedral coordination environments to minimize their crystal field stabilization energy, the early and late transition metals generally have less preference and can therefore occupy a wide variety of coordination environments, such as the 8-coordinated *A* sites.[36] This, in addition to the size effect, whereby many of the intermediate transitional metals are too small to completely fill the *A* sites, contribute to the instability of intermediate transition metals in the *A* site.

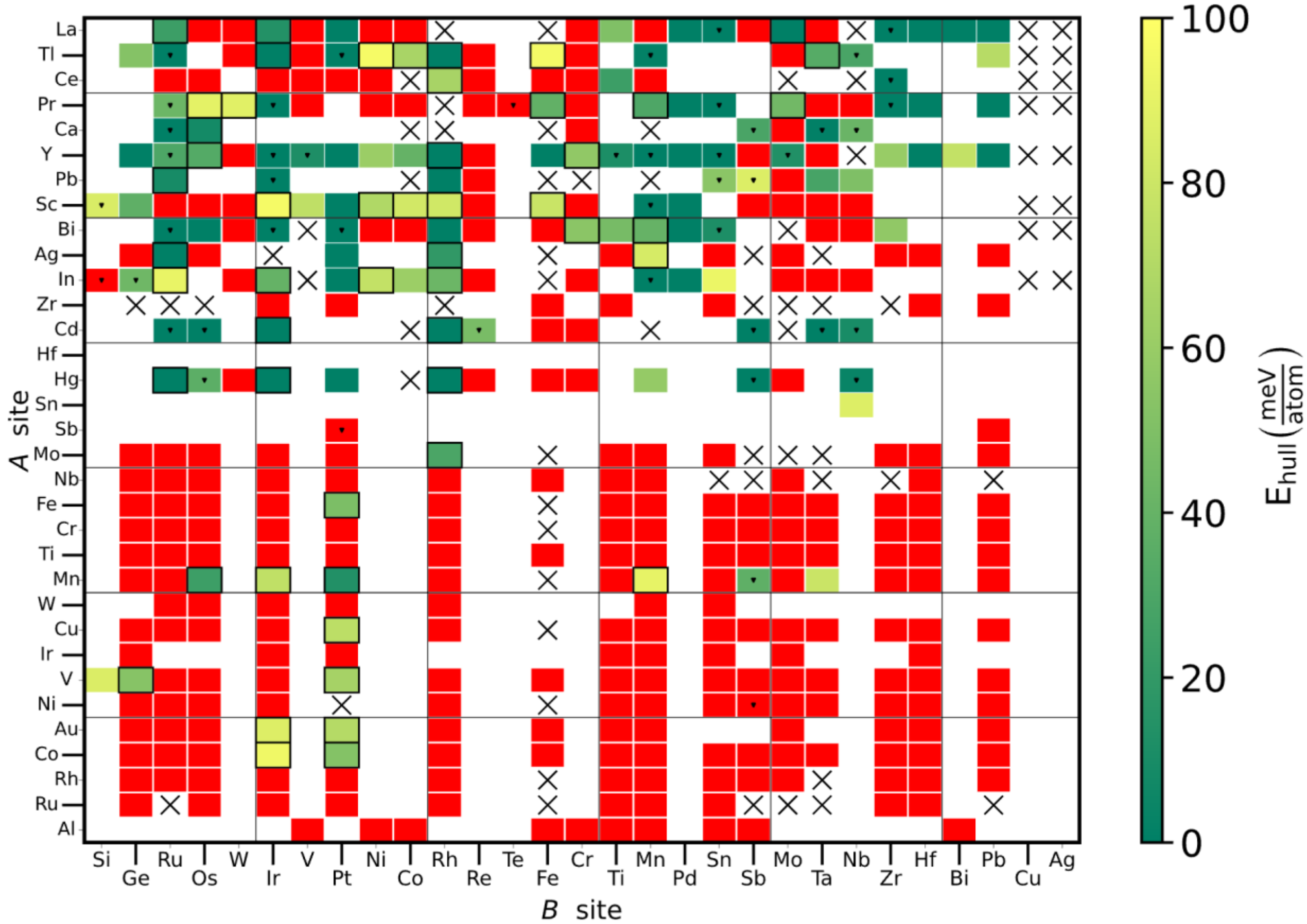

**Figure 3.** Thermodynamic stability map for materials in the pyrochlore structure. Elements are ordered from smallest to largest in terms of their atomic radii. New compounds, which are not present in Materials Project or GNoME, that are also stable or near (< 100 meV/atom) the convex hull are outlined with a black border. Compounds far above (> 100 meV/atom) the convex hull are colored red. Previously synthesized compounds are indicated by ▿. Entries marked by an ✕ were evaluated by DMSP and discarded for low substitution probability. White cells were not pursued. These correspond to compositions that were either lacking unpaired electrons or a viable substitution pathway from a known, stable material.

In **Fig. 4**, we show the relationship between stability and the ionic radii of each element in the *A* and *B* sites of the pyrochlore structure. The shaded region represents pairs of elements that are anticipated to crystallize in the pyrochlore structure based on the ratio of their ionic radii ($1.36 \leq r_A/r_B \leq 2.3$) according to recently proposed heuristics.[25] Interestingly, this heuristic does indeed capture all the experimentally known (*i.e.*, reported in the ICSD) pyrochlores. It is further shown that nearly all (46 of 52) of the new, hypothetical pyrochlores computed to be stable or near the hull fall within the proposed range of ionic radii for the pyrochlore structure. However, there are also many (164 of 263) highly unstable hypothetical pyrochlore materials which would otherwise be predicted as stable based on ionic radii alone. An additional 75 of 148 $A_2B_2O_7$ compositions fall in this range of ionic radii but were experimentally found to crystallize in non-pyrochlore structures. We therefore conclude that while radii-based heuristics provide reasonable bounds to find new pyrochlore materials, they are insufficient to ensure their stability.

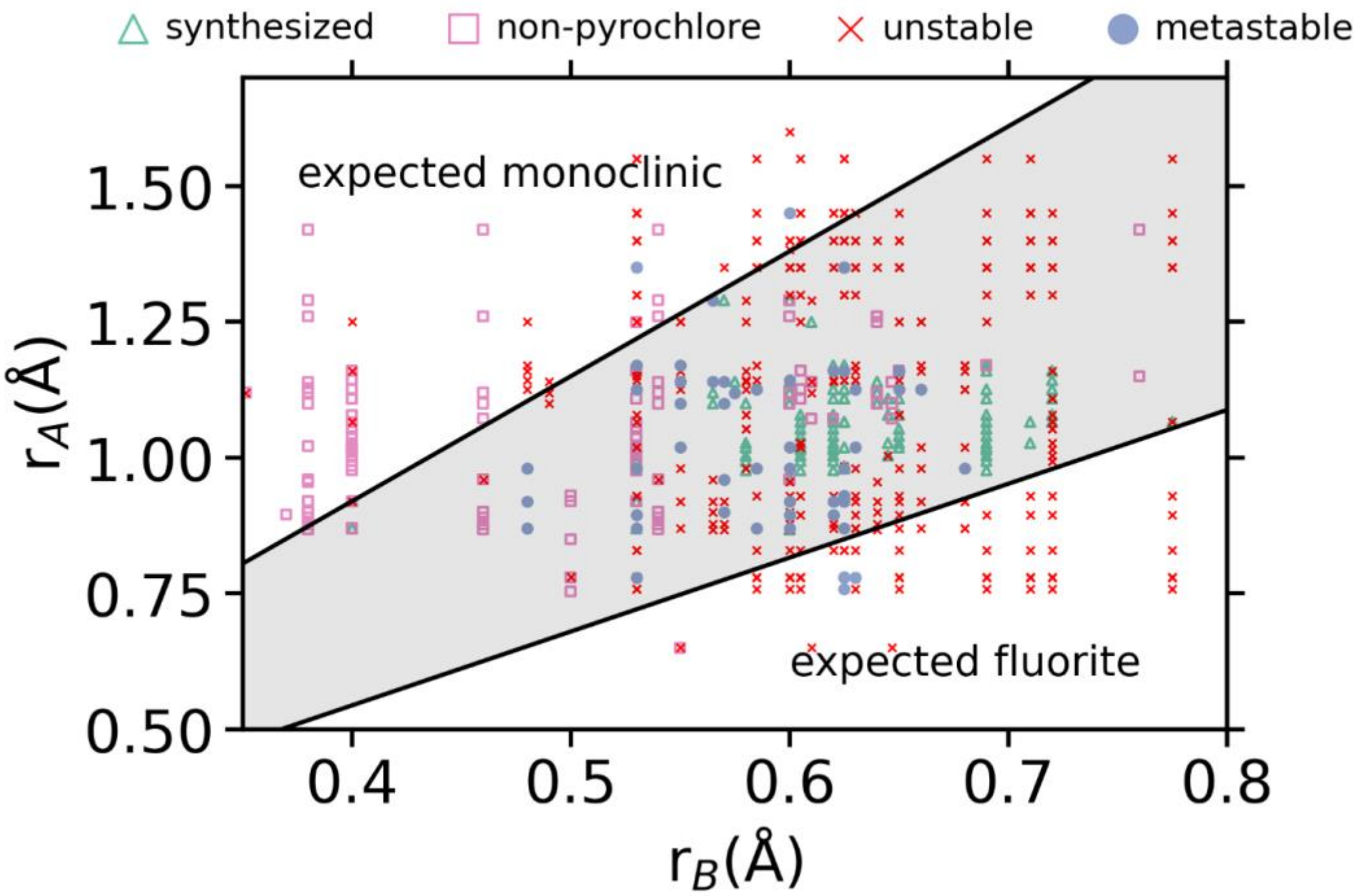

**Figure 4.** Pyrochlore compositions evaluated in this work are plotted in terms of their ionic radii for the elements on the *A* and *B* sites. The range of ionic radii pairs anticipated to stabilize the pyrochlore structure based on previously proposed heuristics[25] is shown by the shaded region ($1.36 \leq r_A/r_B \leq 2.3$). Experimentally known pyrochlore materials are denoted by green triangles, while $A_2B_2O_7$ compositions experimentally observed to crystallize into other structures are denoted by pink squares. Hypothetical materials found to be stable or close (> 100 meV/atom) to the convex hull are denoted by blue circles and those found to be highly unstable are denoted by red crosses. The size of ions in oxidation states and coordination environments without Shannon radii are estimated using an ionic-radius regression model.[37]

While much of the previous work on pyrochlore synthesis has focused on large cations such as 5*d* transition metals and lanthanides to fill the 8-coordinated *A* site, our findings suggest that several emerging families of pyrochlores could make use of relatively smaller cations. Notable pyrochlore series such as $In_2B_2O_7$, $Mn_2B_2O_7$, $Sc_2B_2O_7$, and $Ag_2B_2O_7$ contain many promising compounds that are stable or near the convex hull. We also observe similar trends in the $A_2Os_2O_7$ and $A_2Fe_2O_7$ pyrochlore series, though previous synthesis attempts have shown little success[38,39]. It should be noted that many of the novel pyrochlore materials we report contain unusual oxidation states (*e.g.*, $Ag^{3+}$, $Mo^{3+}$, $Ni^{4+}$, $Cr^{4+}$, $Fe^{4+}$, $Ir^{5+}$) that may be difficult to synthesize using traditional synthesis methods, and precise control over temperature and partial pressures of gaseous species (such as $O_2$) will likely be necessary to achieve these oxidation states.[40–42]

Of the 110 DMSP-generated spinels computed with DFT in this work, we find 14 that are within 100 meV/atom of the convex hull as shown in **Fig. S1**. Most metastable spinel compositions contain an alkali earth (*e.g.*, Mg and Ca) or late transition metal (*e.g.*, Zn and Cd),[35] which can occupy the tetrahedral *A* site without detrimental effects from the crystal field stabilization energy.[36] Similar to the pyrochlore *B* site, the intermediate transition metals have a strong preference to adopt octahedral coordination environments and therefore are more often observed to be stable in the *B* site of the spinel structure. Additionally, we compare our calculations on spinel compositions with the tolerance factor proposed by Song and Liu[26] in **Fig. S2**.[35] While there is some correlation, this tolerance factor is insufficient to describe the instability of compounds such as $MgAg_2O_4$, $CoTi_2O_4$, and $MgMn_2O_4$ which have comparable $r_A$ and $r_B$ to various synthesized spinels but are thermodynamically unstable. Several other $AB_2O_4$ compositions in this radii range are also known to crystallize into non-spinel structures.

Combining the results of our calculations on pyrochlore and spinel materials, DMSP yielded 66 compounds that are thermodynamically stable or within 100 meV/atom of the convex hull. However, only 32 of these compounds showed non-zero magnetic moments after structural and electronic relaxation, despite their initialization in a ferromagnetic configuration. We further evaluated these 32 magnetic compounds by comparing the energy of the ferromagnetic configuration to the energies of three, randomly initialized antiferromagnetic configurations. As an approximation suitable for high-throughput calculations, we limited this analysis to collinear magnetic states and did not consider non-collinear configurations. These calculations reveal that about half (15 of 32) of the compounds exhibit lower energy in at least one of the antiferromagnetic configurations. Therefore, 15 materials remain that are both stable (or close to the hull) and ferromagnetic in their ground state when using the PBE functional. To be more confident in these results, we further evaluated the materials using r$^2$SCAN, a functional that can predict the relative stability of solids more accurately than GGA and GGA+U.[43] These calculations indicate that 13 of these 15 compounds are still predicted to be ferromagnetic in their ground states. We performed this same magnetic sampling approach using r$^2$SCAN for the 60 candidate materials obtained from the Materials Project and GNoME and found only two additional materials that have a ferromagnetic ground state. Combining these sources results in 15 compounds (displayed in **Table 1**) that are considered for further analysis based on thermodynamic stability and magnetic calculations.

To understand the electronic structures of the 15 ferromagnetic materials resulting from our screening efforts, we first evaluated their spin-resolved density of states (DOS) using r$^2$SCAN without spin-orbit coupling (SOC) (**Fig. 5**). Four materials among them ($Mo_2Rh_2O_7$, $V_2Ge_2O_7$, $V_2Pt_2O_7$ and $GeNi_2O_4$) are found to be semiconductors while the rest are metallic. Their magnetic moments range from 1.02 to 4.00 $\mu_B$ per formula unit. In most of these materials, the *d* orbitals from only one of the metallic elements predominantly contributes to the DOS near the Fermi level. For example, the 4*d* orbital in Mo is solely responsible for the electronic structure of $Pr_2Mo_2O_7$ near the Fermi level. As such, it can be well described by the three-dimensional Kagome lattice model.[12,44] However, in materials with Ag ($Ag_2Rh_2O_7$ and $Ag_2Pt_2O_7$), contributions from the *A* and *B* sites are comparable near the Fermi level, leading to interactions between the two Kagome lattices. It is worth noting that in $Pr_2Mo_2O_7$, $Pr_2W_2O_7$, $Tl_2Rh_2O_7$, $In_2Ir_2O_7$, $In_2Rh_2O_7$, $La_2Mo_2O_7$, and $Y_2Rh_2O_7$, only one spin state crosses the Fermi level, indicating half-metallicity. Such intrinsic half-metallic materials are quite rare in nature despite their usefulness.[45] These half-metallic materials may therefore be promising candidates in future spintronics applications.

TABLE 1. Chemical formula, space group, and band gap of the candidates for giant AHE, as calculated using r²SCAN. Energy above hull is reported from initial PBE screening. All of these candidates retain the $Fd\bar{3}m$ space group and were computed to have a ferromagnetic ground state. For semiconducting materials, band gaps were additionally verified using the HSE06 functional, with values shown in parentheses.

| Formula | $E_{hull}$ (meV/atom) | $E_g$ (eV) |
|---|---|---|
| $GeNi_2O_4$ | 0 | 2.14 (3.33) |
| $Ag_2Pt_2O_7$ | 0 | 0 |
| $Cd_2Rh_2O_7$ | 0 | 0 |
| $Hg_2Rh_2O_7$ | 0 | 0 |
| $Tl_2Rh_2O_7$ | 0 | 0 |
| $Y_2Rh_2O_7$ | 0 | 0 |
| $La_2Mo_2O_7$ | 0 | 0 |
| $Ag_2Rh_2O_7$ | 19.2 | 0 |
| $Mo_2Rh_2O_7$ | 29.4 | 1.08 (2.32) |
| $In_2Ir_2O_7$ | 39.6 | 0 |
| $In_2Rh_2O_7$ | 42.4 | 0 |
| $Pr_2Mo_2O_7$ | 44.6 | 0 |
| $V_2Ge_2O_7$ | 52.8 | 1.62 (3.22) |
| $V_2Pt_2O_7$ | 64.9 | 0.81 (2.24) |
| $Pr_2W_2O_7$ | 87.9 | 0 |

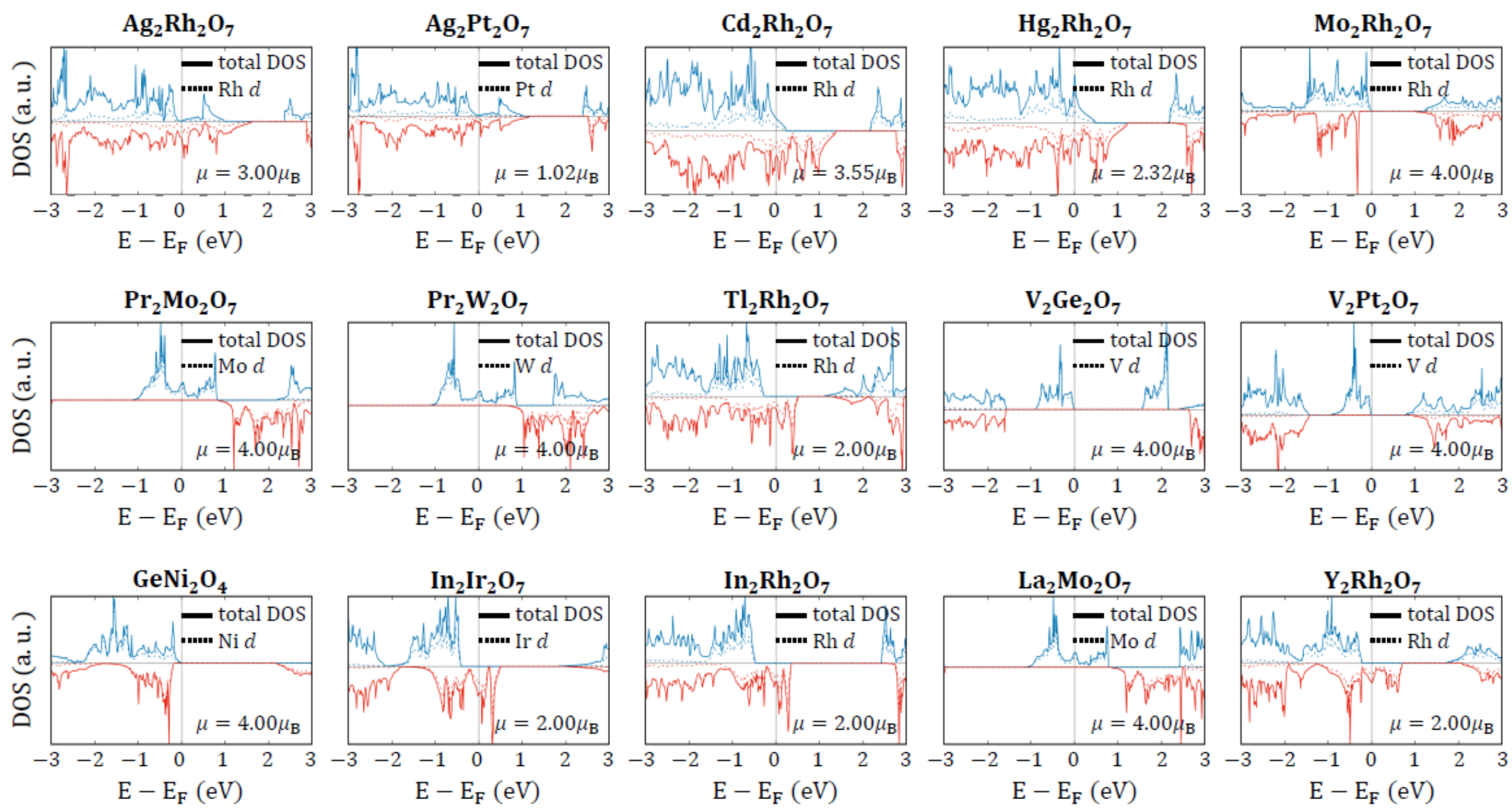

**Figure 5.** Spin-resolved density of states (DOS) of 15 novel ferromagnetic materials (14 pyrochlore and 1 spinel) calculated by the $r^2$SCAN meta-GGA exchange correlation functional without spin-orbit coupling. Blue and red colors represent up and down spins, respectively, and the solid line indicates the total DOS. Dashed lines show contributions of $d$ orbitals from the metallic element that dominantly contributes to the DOS near the Fermi level. Total magnetic moments are also written on each panel in a unit of Bohr magneton per formula unit.

Finally, to evaluate AHC and AHA of the newly proposed materials, we calculated their electronic structures with SOC and ferromagnetic order along the [111] direction. Then, we evaluated AHC and charge conductivity ($\sigma$) by solving the Kubo formula and Boltzmann transport equations, respectively. In **Fig. 6**, we present a summary of the calculated AHC and AHA of our 11 metallic candidates compared to previously reported materials (see **Figs. S3**, **S4** and **S5** for detailed electronic structures, AHC, charge conductivity, and AHA of all 15 materials[35]). To the best of our knowledge, the most promising reported anomalous Hall material is $Co_3Sn_3S_2$, which exhibits high AHC (~1000 S/cm) with a substantial AHA (~0.20).[10] Interestingly, $Co_3Sn_3S_2$ and our materials share similar physics such as Kagome lattice and topological Weyl nodes, suggesting a common underlying physical origin, and potential design rule for anomalous Hall materials. Among our materials, $Ag_2Rh_2O_7$ shows the highest AHC at the intrinsic Fermi level (767 S/cm) with a promising AHA (0.106), which is almost comparable to that of $Co_3Sn_3S_2$. It is worth noting that, as shown in **Fig. S4**, AHC is very sensitive to the position of the Fermi level, $E_F$, while $\sigma$ is relatively insensitive.[35] This is because the former relies on Fermi sea contributions, while the latter is derived from the Fermi surface. This suggests that optimization of the Fermi level can be an effective strategy to enhance both AHC and AHA. Within a reasonable range of experimentally accessible doping ($-0.2 < E_F < 0.2$), either through chemical doping or gating, we found that a high AHA of 0.405 may be achieved in ferromagnetic $Ag_2Pt_2O_7$. We also find three more candidates exhibiting high AHA (at the optimized $E_F$) comparable or greater than that of $Co_3Sn_3S_2$:

$Hg_2Rh_2O_7$ (0.248), $Ag_2Rh_2O_7$ (0.224), and $In_2Rh_2O_7$ (0.211). Note that our predicted AHA can be regarded as a theoretical upper limit because we did not consider any temperature effects (e.g., phonon scattering, Curie temperature) in computing either AHC or $\sigma$.[10] The fundamental physics underlying giant AHC discussed here, however, is not temperature dependent.[12]

Note that $In_2Ir_2O_7$ is also shown in **Fig. 6** to have exceptionally high AHA. However, our prediction of this compound as a ferromagnetic is in conflict with experimental reports that report all-in-all-out (AIAO) antiferromagnetic order.[46] Throughout this work, we rely on the assumption of collinear magnetism, but pyrochlore and spinel compounds often exhibit complex noncollinear magnetic ground states, including AIAO order and even spin liquid phases.[47] However, modeling these configurations in high throughput remains computationally demanding, so our calculations are restricted to collinear magnetism as a means of screening for thermodynamic stability and anomalous Hall response. Nonetheless, it is possible that stabilization of the predicted ferromagnetic phases could be achieved through epitaxial strain[48] or chemical doping[49].

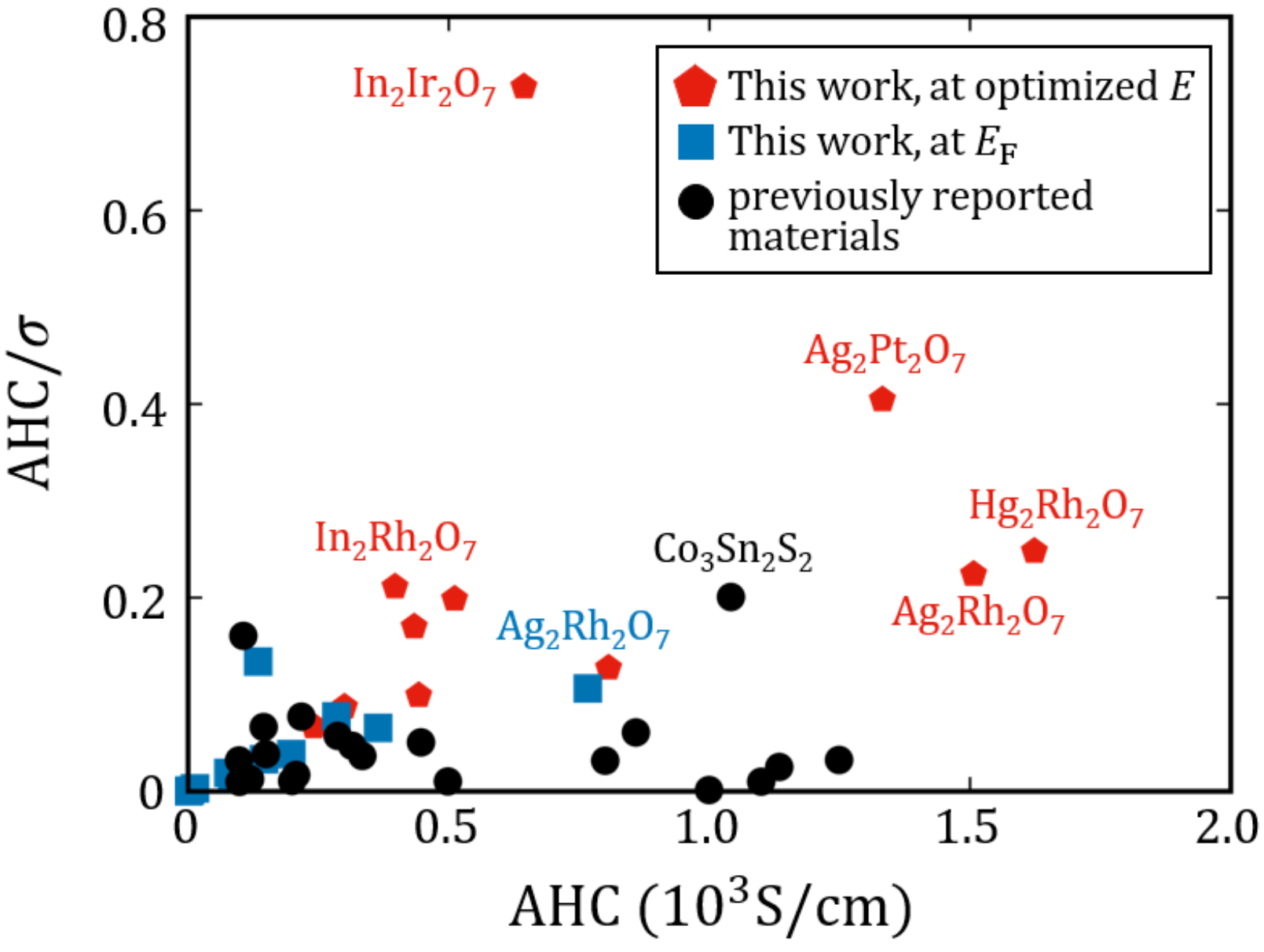

**Figure 6.** Summarized anomalous Hall conductivities (AHC) and anomalous Hall angles (AHC/$\sigma$) of previously reported materials[10] (black), our materials at the computed Fermi level (blue), and our materials at the optimized Fermi level (red) within an energy window of $-0.2 < E_F < 0.2$.

## CONCLUSION

This work presents computational predictions of thermodynamic stability and electronic structure for 448 pyrochlore and spinel materials. Among the materials predicted to be ferromagnetic and stable or near the convex hull, anomalous Hall angle (AHA) and conductivity (AHC) calculations revealed 11 new ferromagnetic metallic materials that are promising candidates for spintronic applications. $Ag_2Rh_2O_7$ exhibits the largest AHA at the intrinsic Fermi level, with a value of 0.106, and doping may be used to further optimize the AHA in other compounds. For example, we predict that $Ag_2Pt_2O_7$ can achieve an exceptional AHA of 0.405 – nearly twice that of $Co_3Sn_3S_2$ – with an appropriate shift in its Fermi level.

Our findings also highlight several emerging families of pyrochlores, extending beyond the traditional focus on 5*d* transition metals and lanthanides. New pyrochlore series including $In_2B_2O_7$, $Mn_2B_2O_7$, $Sc_2B_2O_7$, and $Ag_2B_2O_7$ contain many promising compounds that are stable or near the convex hull. However, experimental investigation is needed to assess the synthesizability of these materials, which will likely require precise control over the synthesis conditions due to the unusual oxidation states (*e.g.*, $Ag^{3+}$, $Mo^{3+}$, $Rh^{5+}$) they contain. More broadly, our analysis demonstrates that existing radii ratio rules for pyrochlore/spinel structures include most stable materials but fail to effectively distinguish between stable and unstable materials. Overall, this work provides a valuable guide for experimentalists, highlighting candidate materials for synthesis and application in next-generation spintronic devices. The identified compounds expand our understanding of the interplay between structure, stability, and electronic properties, offering new pathways for developing high-performance AHE materials.

## METHODS

All DFT calculations were performed using the Vienna *ab initio* Simulation Package (VASP) version 6[50,51] with projector augmented wave pseudopotentials[52,53] to describe the core electrons. Initial screening calculations used the PBE-GGA(+U) exchange correlation functional[32,54], with calculation and convergence parameters selected for structural relaxation following the approach of the Materials Project[28] as described by Jain et al.[55] (using the MPRelaxSet and MPStaticSet in pymatgen)[56]. We adopt fixed U-values from the Materials Project to allow direct comparison of DFT total energies. U-values of 3.32 eV for Co, 3.7 eV for Cr, 5.3 eV for Fe, 3.9 eV for Mn, 4.38 eV for Mo, 6.2 eV for Ni, 3.25 eV for V, and 6.2 eV for W were used. Formation enthalpies computed in this work were compared to formation enthalpies in the Materials Project and GNoME databases using the MaterialsProjectDFTMixingScheme.[57] All calculations were spin-polarized and employed $Fd\bar{3}m$ unit cells except when 2×1×1 supercells are necessary to represent several symmetrically distinct antiferromagnetic initializations. To generate antiferromagnetic initializations, we enumerated through every combination of magnetic sites (initialized as either spin up or spin down) that leads to a net neutral magnetization. We retained symmetrically distinct initializations and randomly selected three of the remaining magnetic initializations. After initial screening using PBE(+U), candidate compounds were furthered investigated using the r$^2$SCAN meta-GGA exchange correlation functional[58]. All calculations were converged to $10^{-6}$ eV/cell for electronic steps and 0.03 eV/Å for ionic steps. A plane wave energy cutoff of 520 eV was used with a Γ-centered Monkhorst-Pack k-point grid and $25|b_i|$ discretizations along each reciprocal lattice vector, $b_i$.

For the selected ferromagnetic materials, we calculated their electronic structures and constructed Wannier Hamiltonians using the r$^2$SCAN meta-GGA exchange correlation functional including spin-orbit coupling without any Hubbard U correction. Then, anomalous Hall conductivity (AHC) was calculated by Kubo-Greenwood formula as implemented in the Wannier90 package[59,60]. For wannierization, $p$ and $d$ orbital projections were used for O and metallic elements, respectively, which guarantees good agreement between first-principles and Wannier electronic structures. To obtain converged AHC, we used a $100 \times 100 \times 100$ k-mesh grid. The longitudinal charge conductivity was calculated by solving the Boltzmann transport equation (BTE) with a constant relaxation time approximation[61]. The fine k-mesh and carrier relaxation time for BTE were chosen to be $80 \times 80 \times 80$ and 10 fs, respectively. The anomalous Hall angle (AHA) is defined as the ratio between AHC and longitudinal charge conductivity.

Throughout our calculations, only $La^{3+}$, $Ce^{3+}$, and $Pr^{3+}$ were included among the lanthanide and actinide elements to mitigate self-interaction errors. To remain consistent with the Materials Project, no Hubbard U corrections were applied to the f-orbitals. Without these corrections, large errors are often observed for elements with partially filled f-shells, particularly mid-series lanthanides.[62] La, Ce, and Pr were still included in our workflow as we anticipate they exhibit lower self-interaction errors relative to other lanthanides with increased f-orbital occupation. DMSP[31] was performed using pymatgen.[56] We started with materials matching our desired candidate criteria ($A_2B_2O_7$ or $AB_2O_4$, $Fd\bar{3}m$, ferromagnetic, $E_{hull} = 0$ meV/atom) from the Materials Project. Then, we considered single isovalent cation substitutions (e.g., substituting Ir in $Ca_2Ru_2O_7$ for Ru to form $Ca_2Ir_2O_7$). Oxidation states were assigned to materials using the oxi_state_guesses method in pymatgen, which determines the most probable oxidation state of ions in a chemical composition based on the ICSD. If the DMSP model predicted a probability > $10^{-5}$ that the new compound is stable (as defined in the DMSP approach), then this new material was computed using PBE(+U). We targeted substitutions that would introduce unpaired $d$-electrons that can contribute to a magnetic moment. Specifically, we considered substitutions of the following ions: $Sc^{2+}$, $Ti^{2+/3+}$, $V^{2+/3+/4+}$, $Cr^{2+/3+/4+/5+}$, $Mn^{2+/3+/4+/5+}$, $Fe^{2+/3+/4+/5+}$, $Co^{2+/3+/4+/5+}$, $Ni^{2+/3+/4+}$, $Cu^{2+/3+/4+}$, $Y^{2+}$, $Zr^{2+/3+}$, $Nb^{2+/3+/4+}$, $Mo^{2+/3+/4+/5+}$, $Ru^{2+/3+/4+/5+}$, $Rh^{2+/3+/4+}$, $Ag^{2+/3+/4+}$, $La^{3+}$, $Ce^{3+}$, $Pr^{3+}$, which were reported to have unpaired $d$ electrons using the get_crystal_field_spin method in pymatgen. Given that the $A$ and $B$ sites in these oxides are coordinated with oxygen, we assumed that cations exhibit characteristics of octahedral and tetrahedral high-spin complexes as expected by crystal field theory to determine ions that should be substituted in. The 8-coordinated pyrochlore $A$ site cations were treated as octahedrally coordinated for electron counting purposes.

## ACKNOWLEDGMENTS

This work was supported by the MRSEC Program of the National Science Foundation under award number DMR-2011401. The authors acknowledge the Minnesota Supercomputing Institute (MSI) at the University of Minnesota (UMN) for providing resources that contributed to the research results reported within this paper. S.S. acknowledges additional support provided by the UMN Undergraduate Research Opportunities Program (UROP).

## DATA AVAILABILITY STATEMENT

The data that support the findings of this article are openly available in digital form as part of the Supplementary Information.

## SUPPLEMENTAL INFORMATION

# Computational search for materials having a giant anomalous Hall effect in the pyrochlore and spinel crystal structures

Sean Sullivan[1^], Seungjun Lee[2^], Nathan J. Szymanski[1], Amil Merchant[3], Ekin Dogus Cubuk[4], Tony Low[2*], and Christopher J. Bartel[1*]

[^]These authors contributed equally
[*]Correspondence to tlow@umn.edu or cbartel@umn.edu

[1]Department of Chemical Engineering and Materials Science, University of Minnesota, Minneapolis, Minnesota 55455, USA

[2]Department of Electrical Engineering and Computer Engineering, University of Minnesota, Minneapolis, Minnesota 55455, USA

[3]Google DeepMind

[4]No current affiliation

**Supplemental File 1.** *supplemental_results.json* is provided as an electronic record of the results presented in this work. This .json is a python dictionary that includes the DFT-computed formation energy, energy above hull, band gap, space group, magnetic ordering, and optimized structure along with their “origin” (DMSP, Materials Project, or GNoME) for 508 compounds considered in this work.

**Table S1.** Formula, energy above hull, and space group of DMSP-generated compounds within 100 meV/atom of the convex hull as calculated by PBE(+U). Bolded rows represent the lowest energy ground state out of 3 AFM + 1 FM initialization, while other compounds are the relaxed properties of 1 FM initialization. FiM = ferrimagnetic; NM = nonmagnetic; FM = ferromagnetic; AFM = antiferromagnetic.

| Formula | $E_{hull}$ (meV/atom) | Space group number | Magnetic Ordering | Eg (eV) |
|---|---|---|---|---|
| **$Cd_2Ir_2O_7$** | **0** | **227** | **FiM** | **0** |
| **$Hg_2Ir_2O_7$** | **0** | **227** | **NM** | **0.04** |
| $Tl_2Ir_2O_7$ | 0 | 227 | NM | 0 |
| **$Cd_2Rh_2O_7$** | **0** | **227** | **FM** | **0.02** |
| $MnGa_2O_4$ | 0 | 166 | FM | 1.46 |
| $Hg_2Ru_2O_7$ | 0 | 227 | NM | 0.06 |
| **$Y_2Rh_2O_7$** | **0** | **227** | **FiM** | **0** |
| **$Hg_2Rh_2O_7$** | **0** | **227** | **FM** | **0.06** |
| **$Tl_2Rh_2O_7$** | **0** | **227** | **FM** | **0.03** |
| **$La_2Mo_2O_7$** | **0** | **227** | **FM** | **0.05** |
| $TiFe_2O_4$ | 0 | 74 | FM | 2.14 |
| $Ag_2Ru_2O_7$ | 0 | 227 | NM | 0.25 |
| **$CdMo_2O_4$** | **0** | **46** | **FiM** | **0.34** |
| $Ca_2Os_2O_7$ | 7.9 | 227 | NM | 0.05 |
| **$Cd_2Mo_2O_7$** | **8.6** | **9** | **FiM** | |
| **$Pb_2Ru_2O_7$** | **9.1** | **227** | **FM** | **0.04** |
| $CdNi_2O_4$ | 10.6 | 227 | FiM | 0 |
| $Mn_2Pt_2O_7$ | 13.5 | 227 | FiM | 0.09 |
| **$La_2Ir_2O_7$** | **13.9** | **227** | **NM** | **0.02** |
| **$Ag_2Rh_2O_7$** | **19.2** | **227** | **FM** | **0.01** |
| **$La_2Ru_2O_7$** | **23.4** | **227** | **FiM** | **0.02** |
| $Mn_2Os_2O_7$ | 23.9 | 227 | FiM | 0.04 |
| $FeCo_2O_4$ | 25.2 | 74 | FiM | 0.96 |
| **$FeMo_2O_4$** | **26.0** | **44** | **FiM** | **0.01** |
| **$Mo_2Rh_2O_7$** | **29.4** | **227** | **FM** | **0.36** |
| **$Bi_2Mo_2O_7$** | **29.6** | **9** | **FiM** | **0.1** |
| $Pr_2Mn_2O_7$ | 30.6 | 227 | FiM | 0.72 |
| **$Y_2Os_2O_7$** | **33.2** | **227** | **FiM** | **0** |
| $Tl_2Ta_2O_7$ | 33.9 | 227 | NM | 0 |
| $CoCu_2O_4$ | 35.3 | 227 | FiM | 0 |
| **$Bi_2V_2O_7$** | **35.7** | **1** | **FiM** | **0.02** |
| **$Pr_2Fe_2O_7$** | **38.4** | **227** | **FiM** | **0.01** |
| **$In_2Ir_2O_7$** | **39.6** | **227** | **FM** | **0.03** |
| $Bi_2Mn_2O_7$ | 40.1 | 227 | FiM | 0.49 |
| **$Ca_2Re_2O_7$** | **41.3** | **166** | **FiM** | **0.01** |
| **$In_2Rh_2O_7$** | **42.4** | **227** | **FM** | **0** |

| **$Pr_2Mo_2O_7$** | **44.6** | **227** | **FM** | **0.05** |
|---|---|---|---|---|
| **$Tl_2Os_2O_7$** | **44.8** | **1** | **FiM** | **0** |
| $HgNi_2O_4$ | 44.9 | 227 | FiM | 0 |
| **$Fe_2Pt_2O_7$** | **48.7** | **227** | **AFM** | **0.30** |
| $FeCu_2O_4$ | 51.3 | 12 | FiM | 0 |
| $Co_2Pt_2O_7$ | 51.6 | 227 | FiM | 0.05 |
| **$V_2Ge_2O_7$** | **52.8** | **227** | **FM** | **0.82** |
| **$VNi_2O_4$** | **54.1** | **227** | **FiM** | **0.01** |
| $Bi_2Cr_2O_7$ | 54.4 | 227 | FiM | 0.02 |
| **$Mo_4O_7$** | **55.2** | **216** | **FiM** | **0.02** |
| $Y_2Cr_2O_7$ | 57.0 | 227 | FiM | 0 |
| **$CrNi_2O_4$** | **58.6** | **227** | **FiM** | **0.19** |
| $HgRh_2O_4$ | 58.9 | 227 | NM | 0.64 |
| $FeMn_2O_4$ | 60.4 | 141 | FM | 1.02 |
| **$V_2Pt_2O_7$** | **64.9** | **227** | **FM** | **0.1** |
| $Ce_2Rh_2O_7$ | 66.0 | 227 | NM | 0.46 |
| **$Tl_2Co_2O_7$** | **66.2** | **227** | **FM** | **0** |
| $BeTi_2O_4$ | 66.9 | 227 | NM | 0.06 |
| $HgMn_2O_4$ | 67.8 | 227 | FiM | 0.02 |
| $AgNi_2O_4$ | 68.3 | 227 | FiM | 0 |
| **$Sc_2Ni_2O_7$** | **69.1** | **227** | **NM** | **0.56** |
| $CdMn_2O_4$ | 69.7 | 227 | FiM | 0 |
| **$In_2V_2O_7$** | **70.1** | **1** | **FM** | **0.02** |
| **$La_2Fe_2O_7$** | **70.7** | **166** | **FiM** | **0** |
| $Au_2Pt_2O_7$ | 70.8 | 227 | NM | 0 |
| $Cu_2Pt_2O_7$ | 73.8 | 227 | NM | 0.02 |
| $MgAu_2O_4$ | 74.7 | 227 | NM | 1.01 |
| $Mn_2Ir_2O_7$ | 75.2 | 227 | FiM | 0 |
| $In_2Ni_2O_7$ | 76.3 | 227 | FiM | 0.43 |
| **$ScNi_2O_4$** | **76.8** | **15** | **FiM** | **0.19** |
| **$Ag_2Ta_2O_7$** | **77.7** | **8** | **NM** | **0.02** |
| **$Cd_2W_2O_7$** | **78.4** | **1** | **FM** | **0** |
| **$Sc_2Fe_2O_7$** | **78.8** | **227** | **FiM** | **0** |
| **$ScFe_2O_4$** | **79.9** | **9** | **FiM** | **0.35** |
| **$In_2Os_2O_7$** | **81.4** | **8** | **FiM** | **0** |
| $AgRh_2O_4$ | 81.5 | 227 | NM | 0 |
| **$Sc_2Rh_2O_7$** | **81.6** | **227** | **FiM** | **0** |
| **$Sc_2Co_2O_7$** | **82.4** | **227** | **FM** | **0** |
| $Ag_2Mn_2O_7$ | 83.9 | 227 | FiM | 0 |
| $BeNb_2O_4$ | 85.9 | 227 | NM | 0 |
| $Au_2Ir_2O_7$ | 86.4 | 227 | NM | 0.02 |
| $BeNi_2O_4$ | 86.6 | 227 | FiM | 0 |
| **$Pr_2W_2O_7$** | **87.9** | **227** | **FM** | **0.05** |
| $Pr_2Os_2O_7$ | 89.3 | 227 | FiM | 0 |
| $Mn_4O_7$ | 90.5 | 227 | FiM | 0 |
| **$In_2Ru_2O_7$** | **91.3** | **227** | **FiM** | **0.02** |
| **$La_2Nb_2O_7$** | **91.6** | **160** | **FiM** | **0** |
| **$CoMn_2O_4$** | **93.5** | **12** | **FiM** | **0.03** |
| $Co_2Ir_2O_7$ | 93.9 | 227 | FiM | 0 |
| **$Ce_2Nb_2O_7$** | **94.3** | **8** | **FiM** | **0** |
| $CrAl_2O_4$ | 94.4 | 74 | FM | 1.69 |
| **$Y_2Nb_2O_7$** | **94.6** | **8** | **NM** | **0.72** |
| **$Zr_2Mo_2O_7$** | **94.7** | **8** | **FiM** | **0.1** |
| **$Tl_2Fe_2O_7$** | **95.5** | **227** | **FiM** | **0** |
| **$Sc_2Ir_2O_7$** | **95.9** | **227** | **FiM** | **0** |
| $Tl_2Ni_2O_7$ | 96.3 | 227 | FiM | 0.17 |

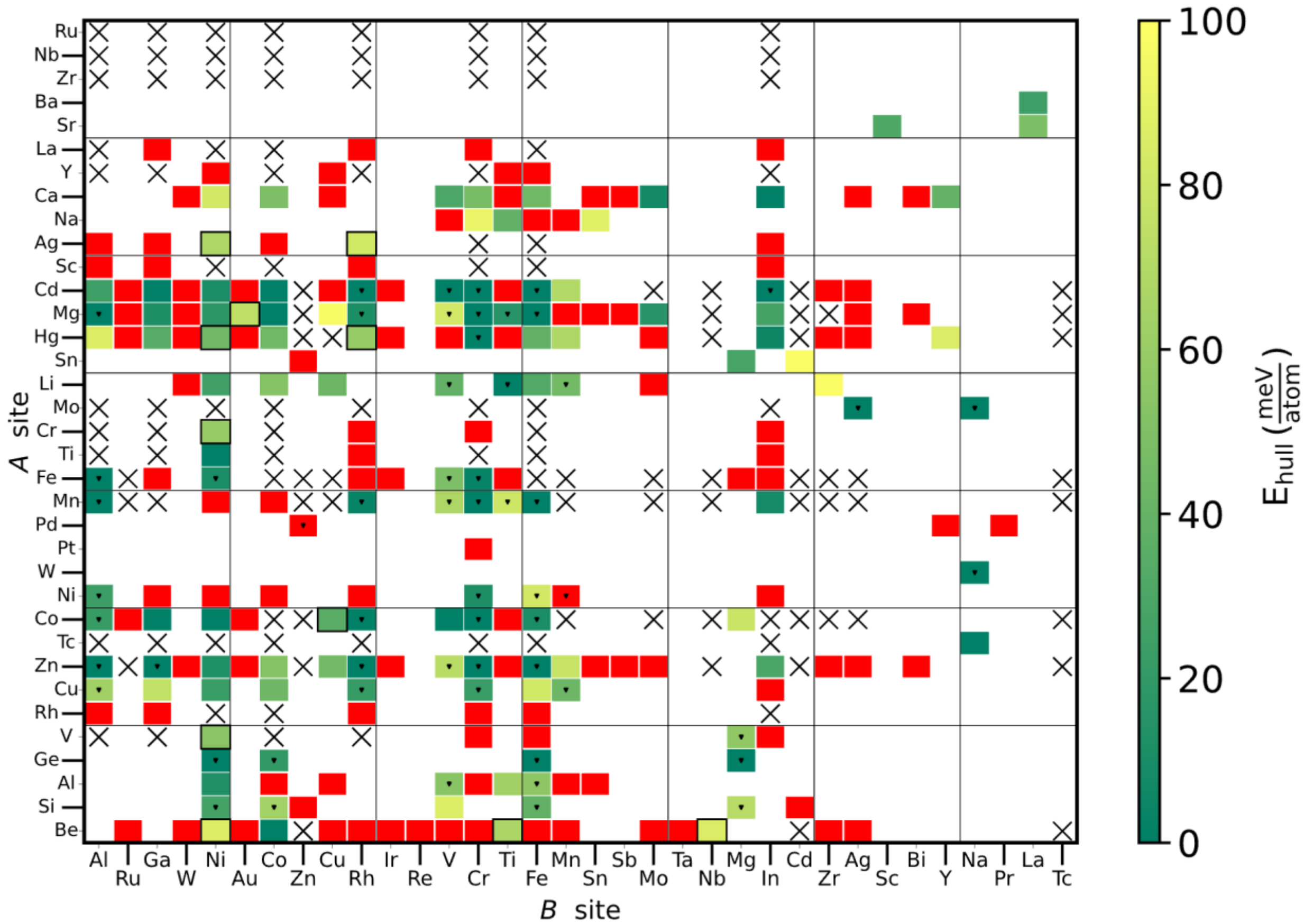

**Figure S1.** Map of thermodynamic stability for materials in the spinel structure. Elements are ordered from smallest to largest in terms of their atomic radii. New compounds, which are not present in Materials Project or GNoME, that are also stable or near (< 100 meV/atom) the convex hull are outlined with a black border. Compounds far (> 100 meV/atom) above the convex hull are colored red. Previously synthesized compounds are indicated by ▽. Entries marked by an X were evaluated by DMSP and discarded for low substitution probability. White cells were not pursued. These correspond to compositions that were either lacking unpaired electrons or were not produced as candidates following ionic substitution into known, stable materials.

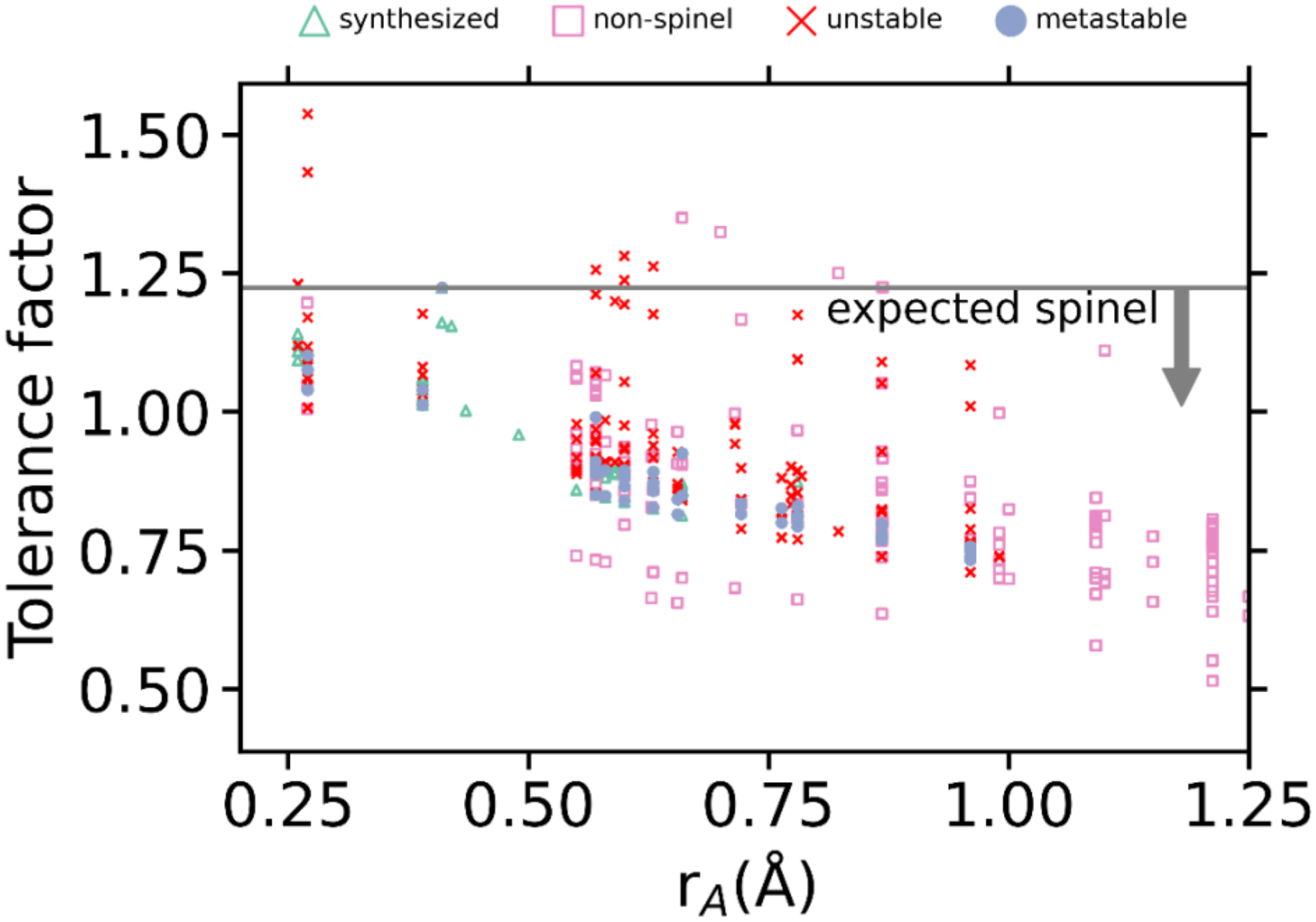

**Figure S2.** Spinel compositions evaluated in this work are plotted in terms of their tolerance factor[26]. Spinel compositions are anticipated to cluster around a tolerance factor of 0.85. Experimentally known spinel materials are denoted by green triangles, while $AB_2O_4$ compositions experimentally observed to crystallize non-$Fd\bar{3}m$ structures are denoted by pink squares. Hypothetical materials found to be stable or close (> 100 meV/atom) to the convex hull are denoted by blue circles and those found to be highly unstable are denoted by red crosses.

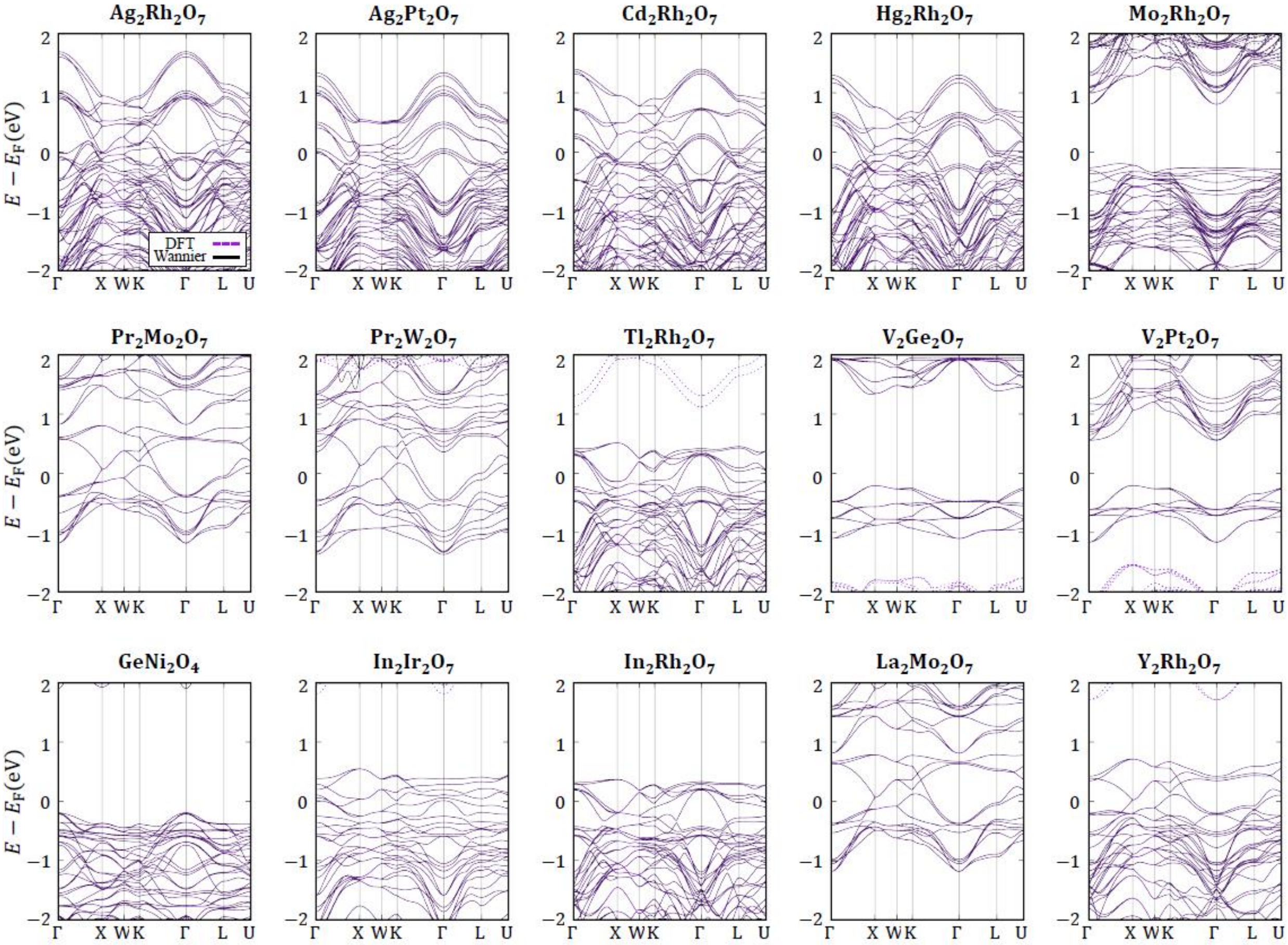

**Figure S3.** Electronic structures of 15 novel ferromagnetic pyrochlore or spinel materials calculated by r$^2$SCAN meta-GGA exchange correlation functional including spin-orbit coupling. Black solid and purple dashed lines represent electronic structures obtained by DFT calculation and Wannier Hamiltonian, respectively.

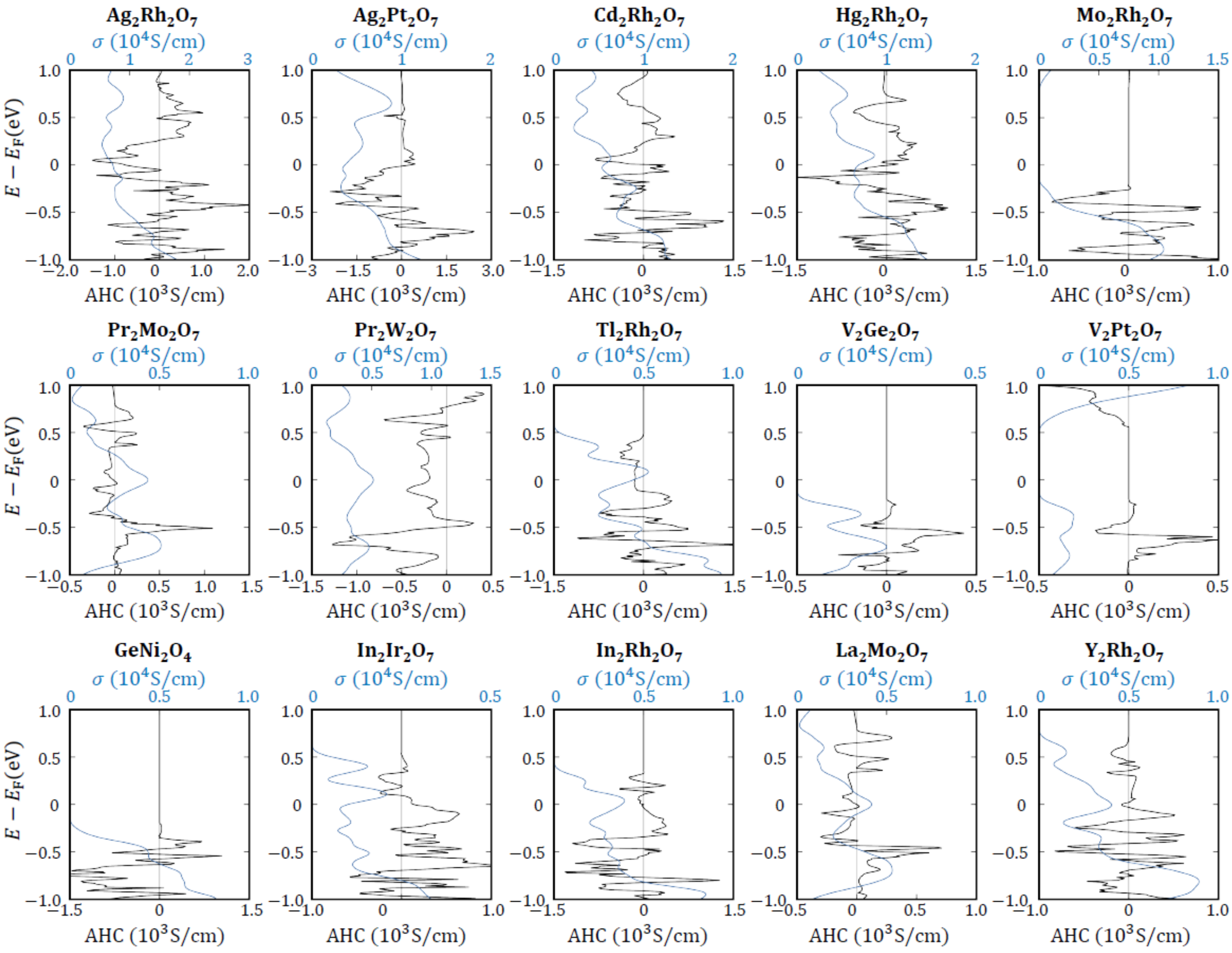

**Figure S4.** Anomalous Hall conductivities (black) and charge conductivities (blue) of 15 novel ferromagnetic pyrochlore or spinel materials.

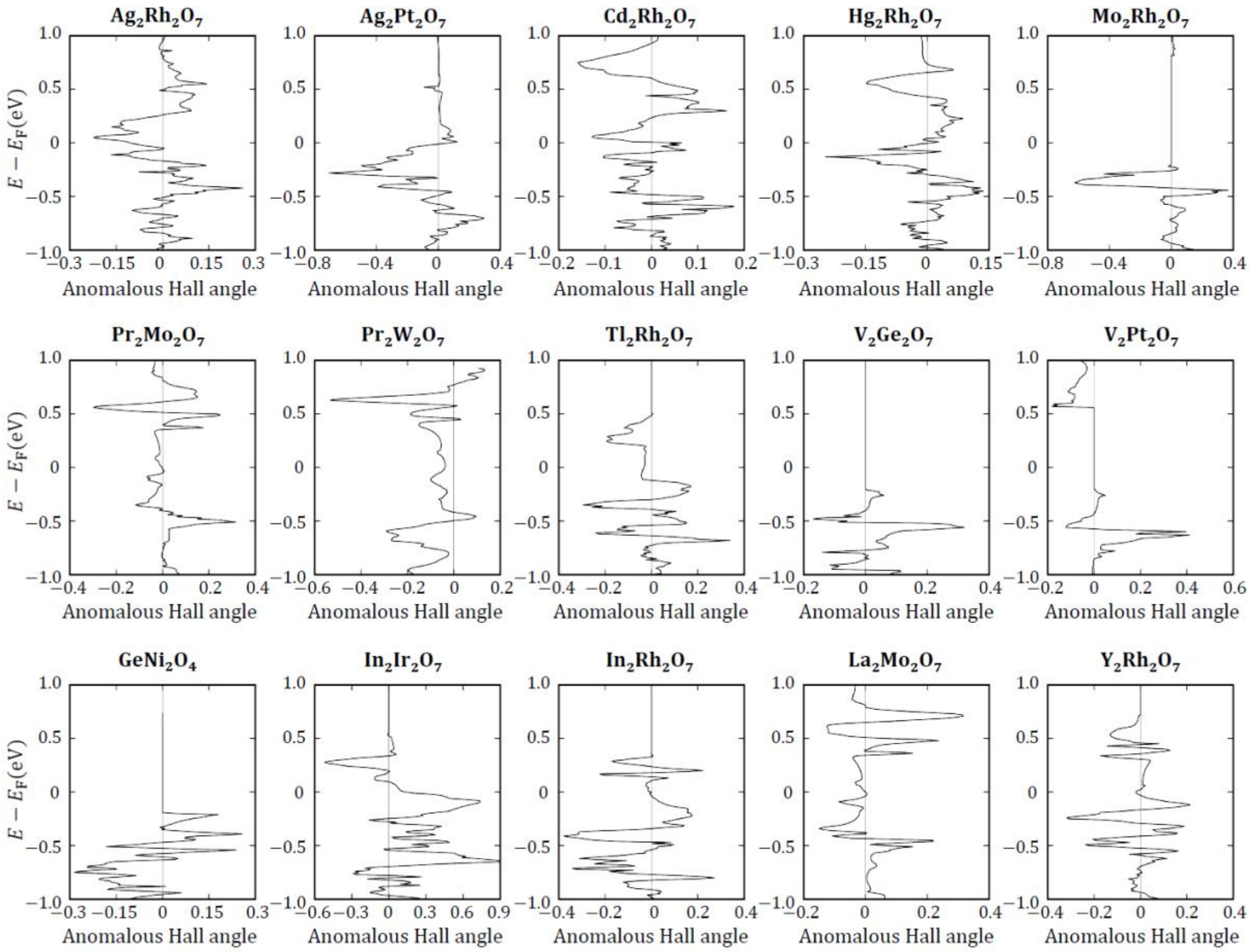

**Figure S5.** Anomalous Hall angles ($\sigma_{xy}/\sigma$ ) of 15 ferromagnetic pyrochlore/spinel materials.